\documentclass{aa}

\usepackage{txfonts}
\usepackage{graphicx}
\usepackage{natbib}

\begin{document}

\title{The flaring and quiescent components of the solar corona}

\author{C.~Argiroffi\inst{1, 2} \and G.~Peres\inst{1, 2} \and S.~Orlando\inst{2} \and F.~Reale\inst{1, 2}}
\offprints{C.~Argiroffi, {\email argi@astropa.unipa.it}}
\institute{Dipartimento di Scienze Fisiche ed Astronomiche, Sezione di Astronomia, Universit\`a di Palermo, Piazza del Parlamento 1, 90134 Palermo, Italy, \email{argi@astropa.unipa.it, peres@astropa.unipa.it, reale@astropa.unipa.it} \and INAF - Osservatorio Astronomico di Palermo, Piazza del Parlamento 1, 90134 Palermo, Italy, \email{orlando@astropa.inaf.it}}

\date{Received 5 January 2008/ Accepted 30 April 2008}

\titlerunning{Flare and quiescent solar coronal contribution}
\authorrunning{C.~Argiroffi et al.}

\abstract
{The solar corona is a template to understand stellar activity. The Sun is a moderately active star, and its corona differs from that of active stars: for instance, active stellar coronae have a double-peaked emission measure distribution $EM(T)$ with the hot peak at $8-20$\,MK, while the non flaring solar corona has one peak at $1-2$\,MK and, typically, much cooler plasma.}
{We study the average contribution of flares to the solar emission measure distribution to investigate indirectly the hypothesis that the hot peak of the $EM(T)$ of active stellar coronae is due to a large number of unresolved solar-like flares, and to infer properties on the flare distribution from nano- to macro-flares.}
{We measure the disk-integrated time-averaged emission measure, $EM_{F}(T)$, of an unbiased sample of solar flares analyzing uninterrupted GOES/XRS light curves over time intervals of one month. We obtain the $EM_{Q}(T)$ of quiescent corona for the same time intervals from the Yohkoh/SXT data. To investigate how $EM_{F}(T)$ and $EM_{Q}(T)$ vary with the solar cycle, we evaluate them at different phases of the cycle between December 1991 and April 1998.}
{Irrespective of the solar cycle phase, $EM_{F}(T)$ appears like a peak of the distribution significantly larger than the values of $EM_{Q}(T)$ for $T\sim5-10$\,MK. As a result the time-averaged $EM(T)$ of the whole solar corona is double-peaked, with the hot peak, due to time-averaged flares, located at temperature similar of that of active stars, but less enhanced. The $EM_{F}(T)$ shape supports the hypothesis that the hot $EM(T)$ peak of active coronae is due to unresolved solar-like flares. If this is the case, quiescent and flare components should follow different scaling laws for increasing stellar activity. In the assumption that the heating of the corona is entirely due to flares, from nano- to macro-flares, then either the flare distribution or the confined plasma response to flares, or both, are bimodal.}
{}


\maketitle

\section{Introduction}
\label{intro}

Late type stars manifest X-ray emission from their coronae: hot plasma ($T\sim1-20$\,MK) magnetically confined in the outer atmosphere. Stellar coronae show a large variety of X-ray luminosity: from $\sim10^{26}-10^{27}\,{\rm erg\,s^{-1}}$ in low activity stars, among which the Sun, to $\sim10^{31}\,{\rm erg\,s^{-1}}$ in very active stars \citep[see][and references therein]{FavataMicela2003,Guedel2004}. The physical differences among the coronae of stars with different activity levels are debated.

To match the higher X-ray luminosities, it was proposed that active stars could have large surface coverage of solar-like coronal structures \citep{WalterCash1980}. However some recent findings have proved that this is not the case. In fact active stars have non-uniform and incomplete surface coverage, like the Sun, as evidenced with: X-ray rotational modulation \citep{GuedelSchmitt1995,MarinoMicela2003}; small coronal volume inferred from measured plasma densities \citep{TestaDrake2004b,NessGuedel2004}, and from resonance scattering \citep{TestaDrake2004a,TestaDrake2007}. Moreover, even a complete surface coverage of solar-like non-flaring coronal structures cannot explain the X-ray emission level of high-activity stars \citep{DrakePeres2000,PeresOrlando2004}.

Active stellar coronae differ from the solar one also for the average plasma temperatures. The emission measure distribution vs. temperature, $EM(T)$\footnote{$EM(T)$, used to characterize optically thin hot plasmas, is defined as the amount of emission measure, $N_{\rm e}^2\,\Delta V$, of plasma with temperature ranging from $T-\Delta T/2$ to $T+\Delta T/2$.} of active stellar coronae usually is peaked around $\sim8-20$\,MK, and, in some cases, with large amount of emission measure at very high temperatures \citep[$T\sim20-30$\,MK,][]{GuedelGuinan1997,SanzForcadaBrickhouse2002,SanzForcadaBrickhouse2003,ScelsiMaggio2005}. The hot bump in the $EM(T)$ of active stars has been observed also in absence of evident flaring activity, i.e. a time independent emission \citep[e.g.][]{SanzForcadaMicela2002,ArgiroffiMaggio2003}.
Conversely on the Sun, and less active stars, the $EM(T)$ peaks at cooler temperatures \citep[$T\sim1-2$\,MK,][]{PeresOrlando2000,RaassenNess2003}.

On the Sun significant amount of emitting plasma at very high temperatures is observed only during flares \citep[e.g.][]{RealePeres2001}. So it has been suggested that the $\sim8-20$\,MK peak in the $EM(T)$ of active stars, could be the result of a large number of unresolved solar-like flares \citep[e.g.][]{GuedelGuinan1997}. However this hypothesis cannot be tested directly because a large number of solar-like flares would produce light curves indistinguishable from a steady emission for the combined effect of flare superposition and available $S/N$ ratio.

Since the hot bump in the $EM(T)$ of active stars may be due to an average process over the stellar disk and over the observing time, in this paper we explore the hypothesis that also the Sun may show such a bump when observed full disk and averaging over a sufficiently long time; we incidentally test also the hypothesis that the $\sim8-20$\,MK peak in the $EM(T)$ of active stars is the result of a large number solar-like flares. To this end we derive the averaged $EM(T)$ of a large number of flares on the solar corona, and compare it with the hot peak of the active coronae $EM(T)$.

This study allows us check whether the reason for the absence of a double peak in the solar $EM(T)$ is that solar flares are rather sparse and that they are observed with different methods and observing modes than the quiescent solar corona and active stars. At the same time with this study we may show a link between the solar and stellar coronae.

The work presented in this paper partly follows the approach of \citet{OrlandoPeres2000} and \citet{PeresOrlando2000}. They analyzed data gathered with the Soft X-ray Telescope \citep[SXT,][]{TsunetaActon1991}, on board the {\it Yohkoh} satellite, to derive the $EM(T)$ of the solar corona including both the quiescent components \citep[quiet corona, active regions, active region cores,][]{OrlandoPeres2001,OrlandoPeres2004}, and individual flares \citep{RealePeres2001}.

The {\it Yohkoh}/SXT normally works in the quiet mode, when the whole solar corona is observed. An abrupt increase of count rate makes it switch to the flare observing mode: only the flaring region is monitored with a high sampling cadence. When the count rate decreases the satellite switches again to the quiet mode. Therefore the flare mode is dedicated to portions of the solar corona for limited time intervals. As a consequence the {\it Yohkoh}/SXT flare observations on the one hand miss information on the whole disk coronal emission, on the other they lack long and continuous time coverage, therefore do not allow an estimation of the average $EM(T)$ of all the solar flares. To fill this gap we analyzed the data gathered with the X-ray Sensor (XRS) photometers \citep{Garcia1994} on board the {\it GOES} satellite. The {\it GOES}/XRS provides continuous disk-integrated flux of the solar corona in two different X-ray bands with a time resolution of $\delta t\sim3$\,s. We devised a key to derive time-averaged flaring $EM(T)$ distributions using this disk-integrated data.

We analyzed {\it GOES}/XRS data over long uninterrupted time intervals, to include a representative and large set of detectable solar flares, and hence to derive an average $EM(T)$ of solar flares. We use this average flaring $EM(T)$ of the Sun to interpret that of active stars in terms of stellar flares. This hypothesis relies on the assumption that the set of phenomena occurring on the Sun are to some extent equivalent to that occurring on more active stars on shorter time scales; thus time-averaging over an adequately long time interval would lead to something similar to the scenario derived from disk-integrated time-averaged stellar observations.

The derivation of the average flaring $EM(T)$, together with the derivation of the quiescent $EM(T)$ \citep{PeresOrlando2000}, allows us to evaluate for the first time also a {\it total} $EM(T)$ of the solar corona, i.e. an $EM(T)$ which includes both the quiescent corona and flares. The estimation of the {\it total} $EM(T)$ is important also for investigating the coronal heating mechanism \citep[e.g.][]{Cargill1994}. \citet{Parker1988} suggested that coronal plasma could be entirely heated by a large number of small magnetic reconnections. This idea relies on the observed frequency distribution of flares vs. energy: many studies showed that $dN/dE\propto E^{-\alpha}$, with $\alpha\sim2$ \citep[e.g.][]{Hudson1991}. The total amount of energy deposited by flares in corona depends on the maximum and minimum flare energy and on the exact value of $\alpha$. Therefore, given reasonable values for these quantities, the high temperature coronal plasmas could be entirely explained by the energy released by flares. For a given model of corona, i.e. a fixed heating function and loop geometry, the temperature structure is entirely constrained (plasma cooling via conduction or radiation is known), hence also the $EM(T)$. Therefore different heating mechanisms predict different $EM(T)$, and the observed $EM(T)$ is a key test for any coronal heating model \citep{Cargill1994}.

In Sect.~\ref{method} we present the method applied to derive the average $EM(T)$ of flaring and quiescent plasma of the solar corona from the {\it GOES}/XRS and {\it Yohkoh}/SXT data, and the cross calibration between these two instruments. The derived results are reported in Sect.~\ref{res} and discussed in Sect.~\ref{disc}.

\section{Method}
\label{method}

Hereafter we indicate with {\it quiescent} corona that observed in the full disk images of {\it Yohkoh}/SXT, i.e. when individual flares are either absent or, if present, very weak, and the total emission is characterized by variations on long time scales (days). On the other hand with {\it flaring} emission we indicate the part of emission due to major flares, i.e. identified in {\it GOES}/XRS light curves.

\begin{figure}
\centering
\includegraphics[width=8.5cm]{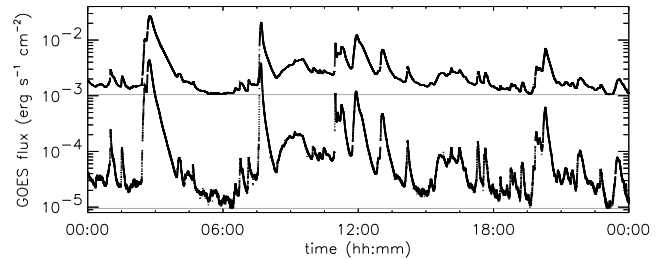} 
\caption{{\it GOES}/XRS light curves in the soft and hard bands registered on December~5~1991. Gray lines mark the minimum emission in the two bands for this 24\,h interval.}
\label{fig:goeslc}
\end{figure}

\subsection{Flaring $EM(T)$}
\label{flareem}

To obtain the flaring $EM(T)$, we analyzed the disk-integrated X-ray light curves recorded by {\it GOES}/XRS, because they grant a complete time coverage of solar flares and a complete sample of events, at variance with {\it Yohkoh}/SXT data-set. We analyzed long time intervals of {\it GOES}/XRS data to recover representative samples of solar flares. We take time intervals of one month because they are:

\begin{itemize}

\item[-] significantly longer than the duration of individual solar flares (from few minutes to few hours);

\item[-] long enough to embrace a large number of flares;

\item[-] long enough to monitor all the solar surface and its magnetic field, and hence to avoid effects due to rotational modulation;

\item[-] short enough not to cancel out the long-term evolution of the solar corona and, at the same time, allow us to repeat the analysis at different phases of the solar cycle to explore variations of the flare component.

\end{itemize}

{\it GOES}/XRS data consist of disk-integrated solar flux gathered in two bands, $1-8$\,\AA~and $0.5-4.0$\,\AA, with a time resolution of $\delta t\sim3$\,s. Since we aimed at deriving only the characteristics of the flaring plasma, we evaluated and subtracted the contribution of quiescent corona to the observed fluxes. For individual flare analysis such contribution is the preflare flux in each band. This may not work for a long time period, because both the soft and hard light curves include the superposition of many flares, especially near the maximum of the solar cycle. Hence we assumed that the quiescent fluxes in the two bands are the minimum values of the fluxes in the 24\,h time interval. Details about the quiescent contribution estimation are given in Appendix~\ref{qflux}. In Fig.~\ref{fig:goeslc} we show, as an example, 24\,h {\it GOES}/XRS light curves whose minimum levels represent the quiescent emission.

For each time bin $\delta t$ we derived the flaring plasma temperature $T$ and the flaring plasma $EM$ applying the flux ratio method \citep[e.g.][]{Garcia1994} to the background subtracted fluxes. This method assumes that the flaring plasma is isothermal in the time $\delta t$. The derived $T$ are an average weighted for the emission measure of all the flaring plasma in that time bin. Hence we obtained a one-month time sequence of $T$ and $EM$ of the total flaring corona.

In order to obtain a time-averaged $EM(T)$ distribution we multiplied each $EM$ value for the duration of its time bin $\delta t$  and divided by the duration $\Delta t$ of the whole month, i.e. $\overline{EM} = EM\,\delta t/\Delta t$. The same approach for the time average, but on a single flare, was applied by \citet{GuedelGuinan1997}.

We considered the same temperature grid adopted for the {\it Yohkoh}/SXT analysis (see Sect.~\ref{quietem}). This grid includes the nominal temperature sensitivity of {\it GOES}/XRS, and its binning is appropriate for the typical temperature accuracy.

Then we summed all the $\overline{EM}$ values whose temperatures fall in the same temperature bin. With this procedure we obtained time-averaged $EM(T)$ of the solar flare component (hereafter indicated as $EM_{F}(T)$).

\subsection{Quiescent $EM(T)$}
\label{quietem}

We derived the quiescent $EM(T)$ of the whole Sun by applying the \citet{PeresOrlando2000} method on the {\it Yohkoh}/SXT images. In the normal observing mode the {\it Yohkoh}/SXT takes images of the whole Sun in different filters. We derived the $EM(T)$ from the images obtained with the Al.1 and AlMg filters. We constructed the $EM$ vs. $T$ histogram of the quiescent corona from the $EM$ and $T$ maps obtained with the filter ratio method \citep{VaianaKrieger1973}, based on the assumption that the plasma comprised in each pixel is isothermal. We adopted the same temperature grid of \citet{PeresOrlando2000}, i.e. a grid ranging from 0.3 to 100\,MK equally spaced in a logarithmic scale. To obtain month-averaged $EM(T)$ (hereafter indicated as $EM_{Q}(T)$) of the slowly changing quiescent solar corona we averaged four $EM(T)$ of the quiescent corona separated by about a week.

{\it Yohkoh}/SXT observations were selected so as to include images of the whole solar corona in two filters (Al.1 and AlMg), with two exposure times (short and long), for a total of four images separated by at most a few minutes \citep[see][for more details]{OrlandoPeres2000}.

\subsection{Cross calibration between Yohkoh/SXT and GOES/XRS data}
\label{crosscal}

We cross-calibrated {\it Yohkoh}/SXT and {\it GOES}/XRS, to compare and use jointly the quiescent and flare emission measure distribution, $EM_{Q}(T)$ and $EM_{F}(T)$, derived from these two instruments. We selected a sample of flares, listed in Table~\ref{tab:flarecal}: a subset of the flares analyzed by \citet{RealePeres2001}\footnote{We discarded the X9 flare of the \citeauthor{RealePeres2001} set because it is oversampled with respect to the other flares, otherwise its weight in the cross-calibration would be too high.}, and the so-called Masuda flare \citep{MasudaKosugi1994}. We analyzed the data on these flares collected with both the instruments. This set embraces a large range of flare intensity, from weak ({\it GOES} class C5.8) to very intense (X1.5), granting an appropriate cross-calibration.

\begin{figure*}
\centering
\includegraphics[width=17cm]{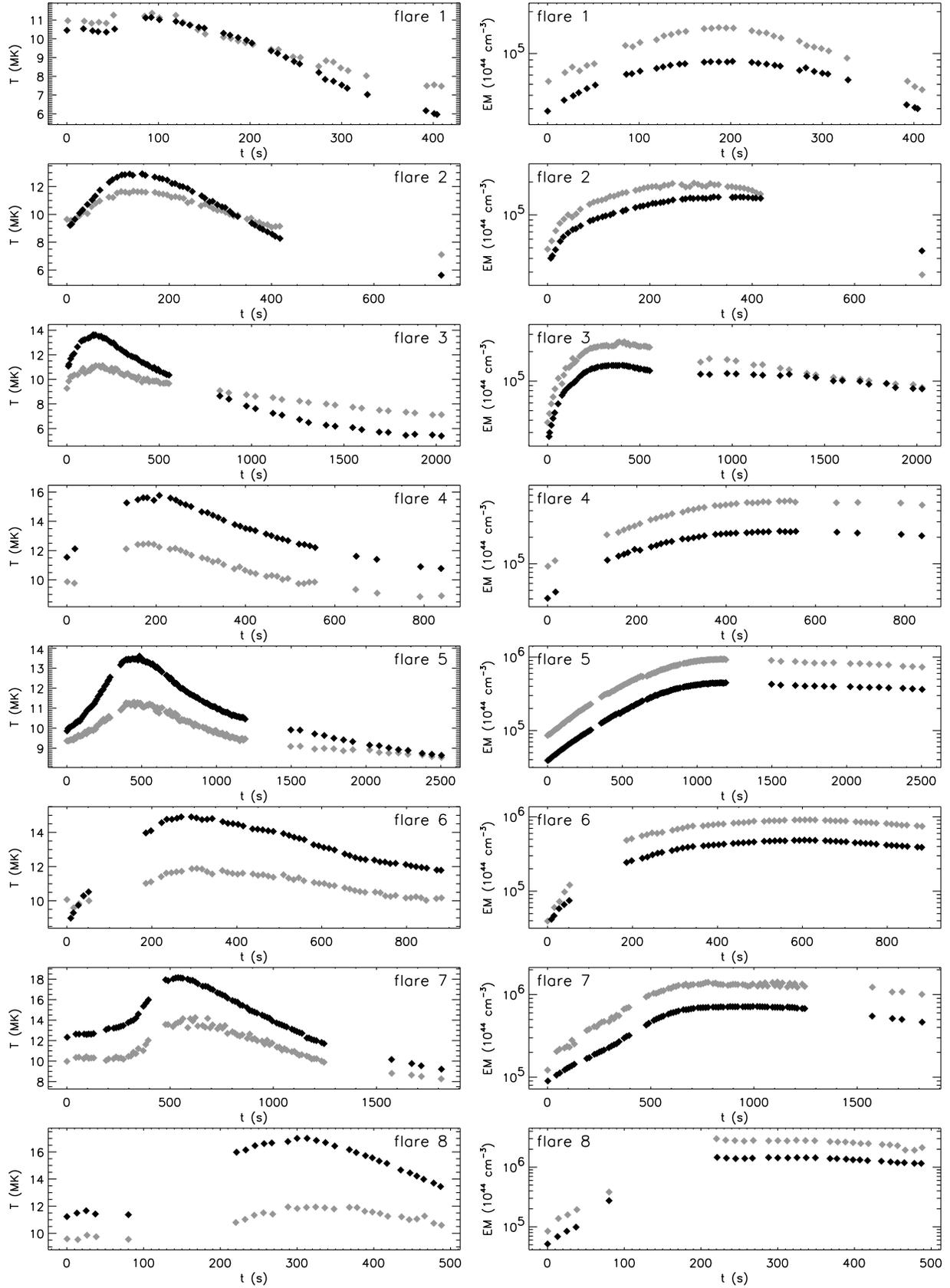} 
\caption{Time series of temperature and total emission measure of the eight flares (see Table~\ref{tab:flarecal}) analyzed for the cross calibration between {\it GOES}/XRS and {\it Yohkoh}/SXT. Black and gray symbols mark the {\it GOES}/XRS and {\it Yohkoh}/SXT data respectively.} 
\label{fig:flarecal}
\end{figure*} 

\begin{figure*}
\centering
\includegraphics[width=8.5cm]{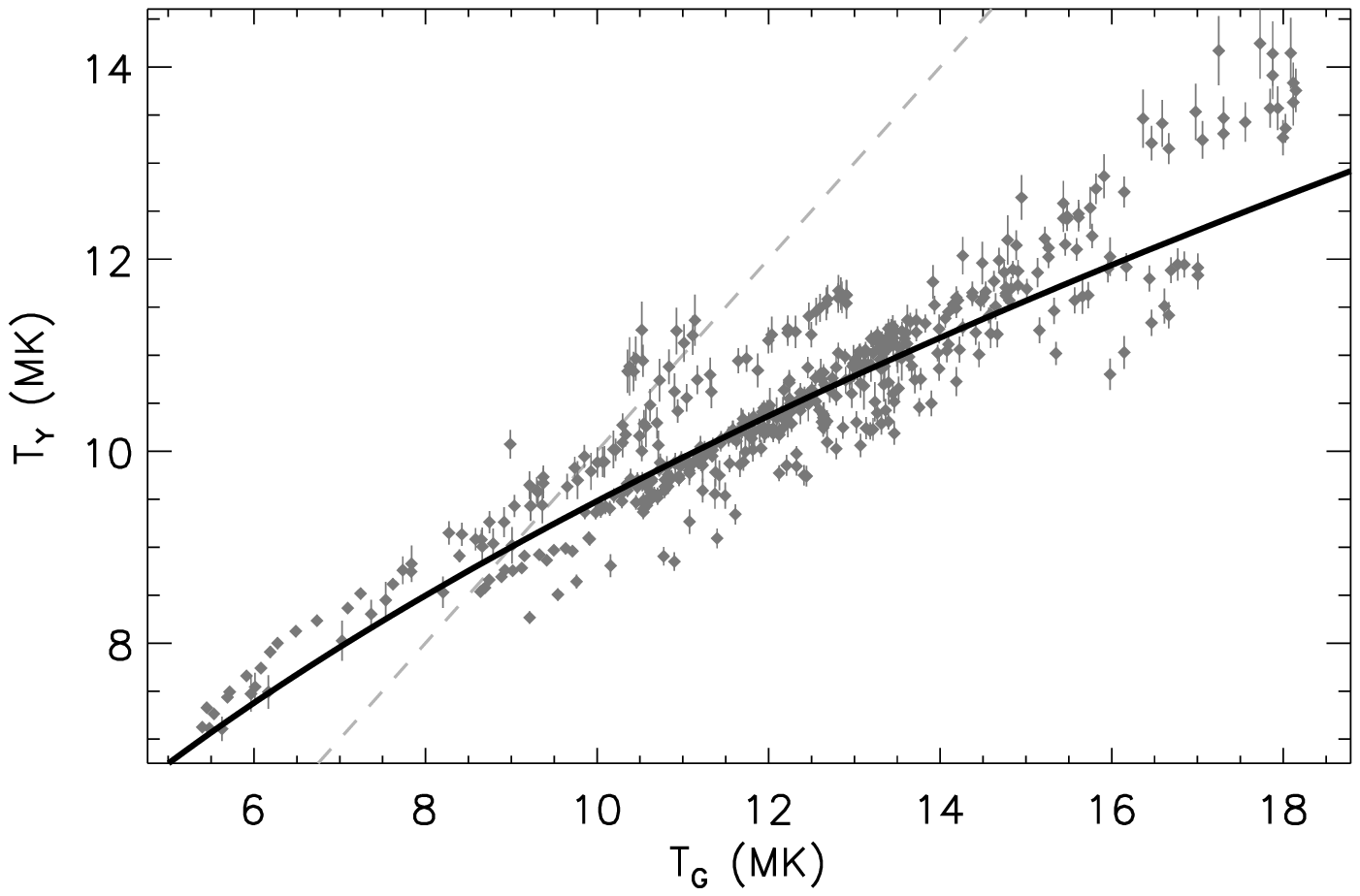} 
\includegraphics[width=8.5cm]{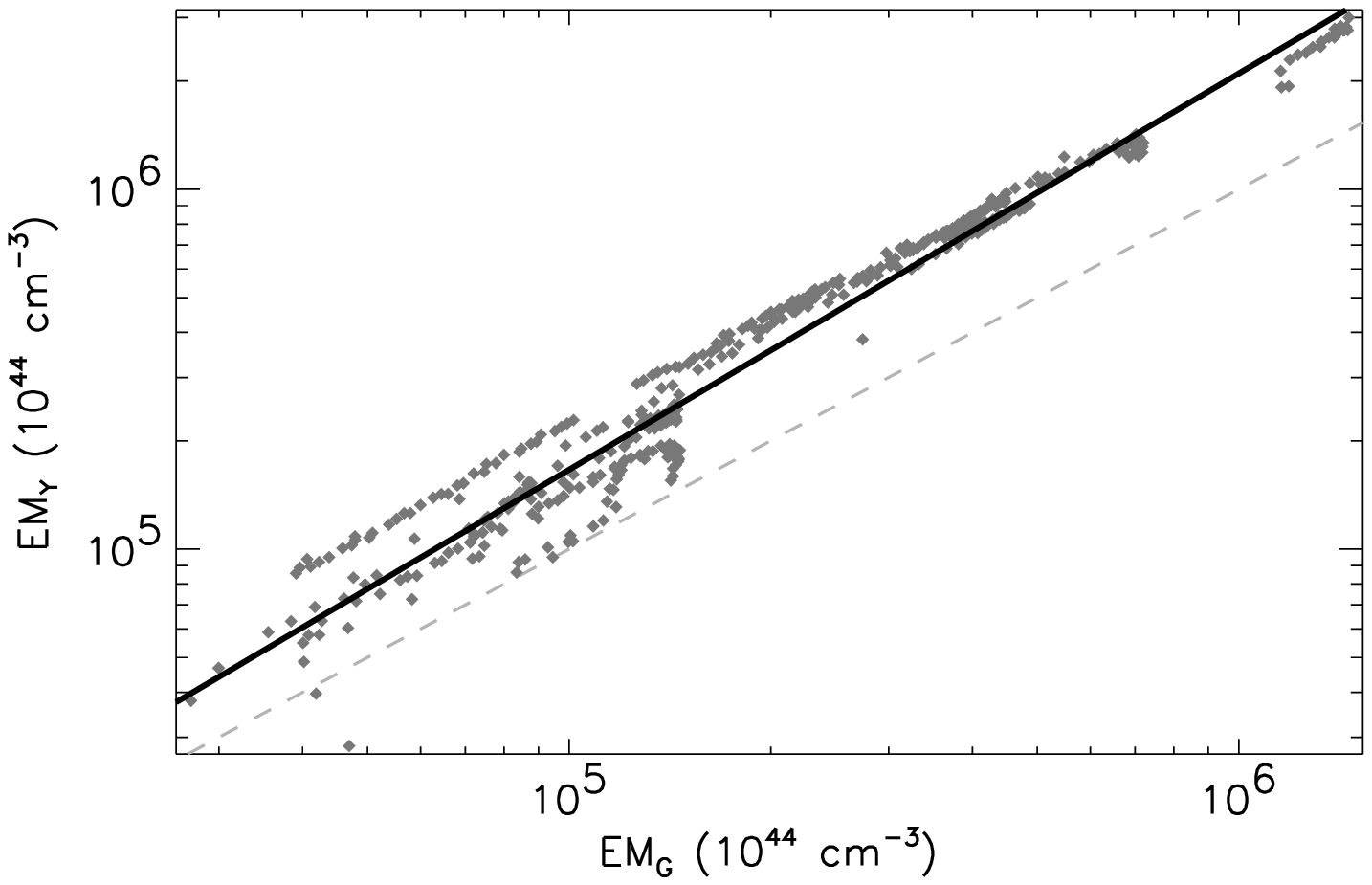} 
\caption{Flare temperatures ({\it left panel}) and emission measure ({\it right panel}) derived from the {\it Yohkoh}/SXT and {\it GOES}/XRS data for the flares listed in Table~\ref{tab:flarecal}. Solid lines indicate the best fit functions, dashed lines the bisectors.} 
\label{fig:crosscal}
\end{figure*} 

We derived the $T$ and $EM$ time sequence from the {\it Yohkoh}/SXT data following the approach of \citet{RealePeres2001}. The {\it Yohkoh}/SXT flares were observed in the flare mode. Each data set consists of a sequence of frames obtained alternatively with the Be~119\,$\mu$m and Al~11.4\,$\mu$m filters, with a time sampling ranging between $\sim10$\,s and $\sim100$\,s. Each frame is composed by images of $64\times64$ pixels of 2\farcs5 side. From each pair of filter images we derived $EM$ and $T$ maps, from that we computed the $EM$-weighted average temperature and the total emission measure.

\begin{table}
\caption{Flares analyzed for the {\it Yohkoh} vs. {\it GOES} cross-calibration.}
\label{tab:flarecal}
\begin{center}
\begin{tabular}{c@{\hspace{2mm}}c@{\hspace{2mm}}c@{\hspace{10mm}}c@{\hspace{2mm}}c@{\hspace{2mm}}c}
\hline\hline
N.      & Date         & Class & N.      & Date         & Class  \\
\hline
1$^a$   & 18~Jul~1992  & C5.8  & 5$^a$   & 27~Feb~1994  & M2.8 \\
2$^a$   & 10~Feb~1993  & M1.0  & 6$^a$   & 26~Dec~1991  & M4.2 \\
3$^a$   & 19~Oct~1992  & M1.1  & 7$^a$   & 06~Feb~1992  & M7.6 \\
4$^b$   & 13~Jan~1992  & M2.0  & 8$^a$   & 15~Nov~1991  & X1.5 \\
\hline
\multicolumn{5}{l}{$^a$~Flares analyzed by \citet{RealePeres2001}. $^b$~Masuda flare.}
\end{tabular}
\end{center}
\end{table}
\normalsize

We derived the simultaneous $EM$ and $T$ values from the {\it GOES}/XRS data as described in Sect.~\ref{flareem}. We subtracted the flux values in the two bands just before the flare start, assuming that during the flare evolution the background coronal emission remains constant. From the {\it GOES}/XRS flare analysis we obtained time series of $T$ and $EM$. Since the time sampling of {\it Yohkoh}/SXT is less frequent than that of {\it GOES}/XRS, we considered $EM$ and $T$ from {\it GOES}/XRS at the times of the {\it Yohkoh}/SXT data.

The derived results of $T$ and $EM$ evolution for each flare from {\it Yohkoh}/SXT and {\it GOES}/XRS are shown in Fig.~\ref{fig:flarecal}. The {\it GOES}/XRS temperatures tend to be higher than those measured with {\it Yohkoh}/SXT, especially for the more energetic flares. The opposite occurs for the emission measure. Figure~\ref{fig:crosscal} shows the correspondence between the values of temperature and emission measure derived from {\it GOES}/XRS and {\it Yohkoh}/SXT, indicated as $T_{G}$, $T_{Y}$, $EM_{G}$, $EM_{Y}$. For the cross-calibration of temperature and emission measure, we assumed a power-law function. The best fit functions are:

\[
T_{Y} = a_{1}T_{G}^{a_{2}}\;\;\;EM_{Y} = a_{3}EM_{G}^{a_{4}} \\
\]

\noindent
with $ a_{1}=3.06\pm0.01$, $a_{2}=0.491\pm0.002$, $a_{3}=0.520\pm0.002$,  $a_{4}=1.1010\pm0.0003$, with temperatures in MK, and emission measure values in $10^{44}\,{\rm cm^{-3}}$. The best fit functions, reproducing well the observed pattern, were used to correct $T$ and $EM$ values derived from the {\it GOES}/XRS data.

The scatter between $T_{Y}$ values, corresponding to similar $T_{G}$ values, is significantly larger than the typical $T_{Y}$ error bar (see Fig.~\ref{fig:crosscal}). It implies that the uncertainty on the inferred $T_{Y}$ values is provided by the scatter of the observed points. The same happens for the emission measure. The observed scatter is 5\% (1$\sigma$) for $T_{Y}$ and 20\% (1$\sigma$) for $EM_{Y}$. The bin widths of the temperature grid over which we constructed the emission measure distributions are comparable to the uncertainties on the derived temperatures.

\begin{table}
\renewcommand{\baselinestretch}{1.05}
\caption{{\it Yohkoh}/SXT and {\it GOES}/XRS data.}
\label{tab:log}
\begin{center}
\scriptsize
\begin{tabular}{l@{\hspace{2mm}}lc@{/}c@{/}c@{\hspace{2mm}}cc@{/}c@{/}c@{\hspace{2mm}}cc@{/}c@{/}c@{\hspace{2mm}}c}
\hline\hline
 \multicolumn{2}{c}{        } & \multicolumn{4}{c}{{\it GOES}/XRS}      & \multicolumn{8}{c}{{\it Yohkoh}/SXT}    \\
 \multicolumn{2}{c}{Analyzed} & \multicolumn{4}{c}{start and stop date} & \multicolumn{8}{c}{obs. 1, 2, 3, and 4} \\
 \multicolumn{2}{c}{month   } & \multicolumn{4}{c}{(dd/mm/yy hh:mm)}    & \multicolumn{8}{c}{(dd/mm/yy hh:mm)}    \\
\hline
 Dec. & 1991          & start: 01 & 12 & 91 & 00:00 & 03 & 12 & 91 & 01:35 & 09 & 12 & 91 & 07:25 \\
\multicolumn{2}{c}{ } & stop:  01 & 01 & 92 & 00:00 & 16 & 12 & 91 & 07:15 & 23 & 12 & 91 & 23:13 \\
\hline
 Apr. & 1992          & start: 01 & 04 & 92 & 00:00 & 04 & 04 & 92 & 01:16 & 11 & 04 & 92 & 03:54 \\
\multicolumn{2}{c}{ } & stop:  01 & 05 & 92 & 00:00 & 18 & 04 & 92 & 03:21 & 25 & 04 & 92 & 23:53 \\
\hline
 Aug. & 1992          & start: 01 & 08 & 92 & 00:00 & 05 & 08 & 92 & 01:34 & 12 & 08 & 92 & 23:35 \\
\multicolumn{2}{c}{ } & stop:  01 & 09 & 92 & 00:00 & 19 & 08 & 92 & 01:46 & 26 & 08 & 92 & 01:04 \\
\hline
 Dec. & 1992          & start: 01 & 12 & 92 & 00:00 & 05 & 12 & 92 & 01:49 & 12 & 12 & 92 & 04:19 \\
\multicolumn{2}{c}{ } & stop:  01 & 01 & 93 & 00:00 & 19 & 12 & 92 & 00:16 & 28 & 12 & 92 & 14:47 \\
\hline
 Apr. & 1993          & start: 01 & 04 & 93 & 00:00 & 05 & 04 & 93 & 01:41 & 14 & 04 & 93 & 21:01 \\
\multicolumn{2}{c}{ } & stop:  01 & 05 & 93 & 00:00 & 19 & 04 & 93 & 01:37 & 26 & 04 & 93 & 00:44 \\
\hline
 Aug. & 1993          & start: 01 & 08 & 93 & 00:00 & 05 & 08 & 93 & 03:56 & 12 & 08 & 93 & 04:39 \\
\multicolumn{2}{c}{ } & stop:  01 & 09 & 93 & 00:00 & 19 & 08 & 93 & 00:29 & 26 & 08 & 93 & 09:17 \\
\hline
 Dec. & 1993          & start: 01 & 12 & 93 & 00:00 & 05 & 12 & 93 & 15:17 & 12 & 12 & 93 & 12:41 \\
\multicolumn{2}{c}{ } & stop:  01 & 01 & 94 & 00:00 & 19 & 12 & 93 & 05:15 & 26 & 12 & 93 & 02:38 \\
\hline
 Apr. & 1994          & start: 01 & 04 & 94 & 00:00 & 04 & 04 & 94 & 23:37 & 12 & 04 & 94 & 01:51 \\
\multicolumn{2}{c}{ } & stop:  01 & 05 & 94 & 00:00 & 18 & 04 & 94 & 23:14 & 25 & 04 & 94 & 23:52 \\
\hline
 Aug. & 1994          & start: 01 & 08 & 94 & 00:00 & 05 & 08 & 94 & 03:04 & 11 & 08 & 94 & 22:46 \\
\multicolumn{2}{c}{ } & stop:  01 & 09 & 94 & 00:00 & 19 & 08 & 94 & 17:11 & 26 & 08 & 94 & 04:54 \\
\hline
 Dec. & 1994          & start: 05 & 12 & 94 & 00:00 & 07 & 12 & 94 & 21:23 & 15 & 12 & 94 & 23:52 \\
\multicolumn{2}{c}{ } & stop:  01 & 01 & 95 & 00:00 & 20 & 12 & 94 & 02:43 & 27 & 12 & 94 & 17:56 \\
\hline
 Apr. & 1995          & start: 01 & 04 & 95 & 00:00 & 04 & 04 & 95 & 06:37 & 11 & 04 & 95 & 15:18 \\
\multicolumn{2}{c}{ } & stop:  01 & 05 & 95 & 00:00 & 18 & 04 & 95 & 12:36 & 24 & 04 & 95 & 01:28 \\
\hline
 Aug. & 1995          & start: 01 & 08 & 95 & 00:00 & 06 & 08 & 95 & 12:42 & 13 & 08 & 95 & 05:13 \\
\multicolumn{2}{c}{ } & stop:  01 & 09 & 95 & 00:00 & 19 & 08 & 95 & 02:16 & 25 & 08 & 95 & 23:34 \\
\hline
 Dec. & 1995          & start: 01 & 12 & 95 & 00:00 & 05 & 12 & 95 & 07:58 & 12 & 12 & 95 & 13:18 \\
\multicolumn{2}{c}{ } & stop:  31 & 12 & 95 & 00:00 & 19 & 12 & 95 & 20:27 & 25 & 12 & 95 & 23:59 \\
\hline
 Apr. & 1996          & start: 01 & 04 & 96 & 00:00 & 05 & 04 & 96 & 01:45 & 12 & 04 & 96 & 11:56 \\
\multicolumn{2}{c}{ } & stop:  01 & 05 & 96 & 00:00 & 19 & 04 & 96 & 12:27 & 24 & 04 & 96 & 23:45 \\
\hline
 Aug. & 1996          & start: 01 & 08 & 96 & 00:00 & 06 & 08 & 96 & 13:22 & 14 & 08 & 96 & 10:56 \\
\multicolumn{2}{c}{ } & stop:  01 & 09 & 96 & 00:00 & 21 & 08 & 96 & 21:08 & 29 & 08 & 96 & 10:36 \\
\hline
 Dec. & 1996          & start: 01 & 12 & 96 & 00:00 & 03 & 12 & 96 & 18:32 & 13 & 12 & 96 & 02:15 \\
\multicolumn{2}{c}{ } & stop:  31 & 12 & 96 & 00:00 & 20 & 12 & 96 & 14:04 & 27 & 12 & 96 & 03:12 \\
\hline
 Apr. & 1997          & start: 01 & 04 & 97 & 00:00 & 04 & 04 & 97 & 00:14 & 11 & 04 & 97 & 07:14 \\
\multicolumn{2}{c}{ } & stop:  30 & 04 & 97 & 00:00 & 18 & 04 & 97 & 15:51 & 25 & 04 & 97 & 03:21 \\
\hline
 Aug. & 1997          & start: 01 & 08 & 97 & 00:00 & 05 & 08 & 97 & 17:38 & 11 & 08 & 97 & 22:38 \\
\multicolumn{2}{c}{ } & stop:  01 & 09 & 97 & 00:00 & 18 & 08 & 97 & 08:31 & 25 & 08 & 97 & 00:51 \\
\hline
 Dec. & 1997          & start: 01 & 12 & 97 & 00:00 & 05 & 12 & 97 & 00:22 & 12 & 12 & 97 & 07:15 \\
\multicolumn{2}{c}{ } & stop:  01 & 01 & 98 & 00:00 & 19 & 12 & 97 & 17:23 & 26 & 12 & 97 & 06:27 \\
\hline
 Apr. & 1998          & start: 01 & 04 & 98 & 00:00 & 05 & 04 & 98 & 22:41 & 13 & 04 & 98 & 13:39 \\
\multicolumn{2}{c}{ } & stop:  01 & 05 & 98 & 00:00 & 18 & 04 & 98 & 00:30 & 25 & 04 & 98 & 17:06 \\
\hline
\end{tabular}
\normalsize
\end{center}
\end{table}

\begin{figure*}
\centering
\includegraphics[width=17cm]{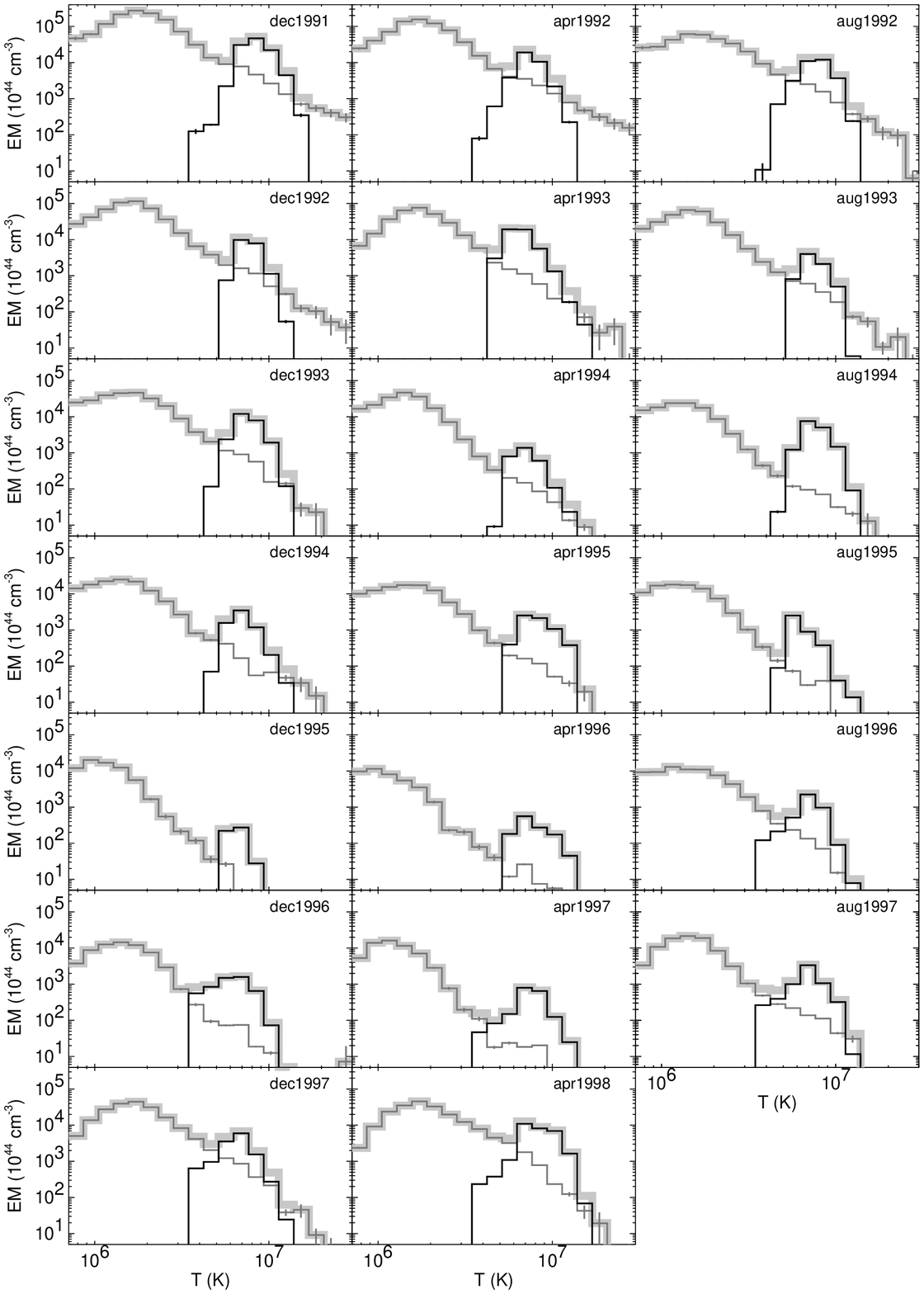} 
\caption{Emission measure distribution vs. temperature of the quiescent (thin line, dark gray) and flaring (thin line, black) solar corona in different months, and their sum (thick line, light gray). Error bars are plotted for each bin, but they are large enough to be visible only for some of the bins.}
\label{fig:fqemd}
\end{figure*} 

\section{Data sets and results}
\label{res}

We derived the emission measure distributions, $EM_{Q}(T)$ and $EM_{F}(T)$, of the quiescent and flaring plasma of the solar corona averaged over one month. Since solar coronal characteristics (i.e. total emission measure, average temperature, flare intensity and frequency) change considerably during the solar cycle, we repeated the derivation of $EM_{Q}(T)$ and $EM_{F}(T)$ at different moments of the solar cycle to probe their dependence on the solar activity level.

We measured the flaring and quiescent $EM(T)$ every four months, obtaining a total of 20 analyzed months between December 1991 and April 1998, i.e. including both the maximum and the minimum phases of the solar cycle. All the dates of the {\it Yohkoh}/SXT and {\it GOES}/XRS data of each analyzed month are listed in Table~\ref{tab:log}. Figure~\ref{fig:fqemd} shows the $EM_{Q}(T)$ and $EM_{F}(T)$, derived for all the inspected months.


The quiescent component - whose shape and variations with the solar cycle were studied by \citet{PeresOrlando2000} - displays a peak at $T\sim1-2$\,MK, and a gradual decrease at higher temperature. $EM_{Q}(T)$ changes during the solar cycle: near the minimum it is lower than near the maximum, its peak is located at cooler temperatures, and its slope at high temperature is steeper.


The first noticeable result is that the $EM(T)$ of the flare component is narrow and peaks at $T\sim6-8$\,MK. This peak is significantly higher than the level of the quiescent plasma at the same temperature range, and it holds irrespective of the phase of the solar cycle. The $EM_{F}(T)$ height follows the solar cycle: it is higher at the solar maximum than at the minimum by a factor $\sim10^{2}$. On the contrary both the peak temperature and the $EM$-averaged temperature of the flare component do not show any clear trend with the phase of the solar cycle. Hence the amount of flaring emission measure varies largely across the solar cycle, while its average temperature does not.


By summing $EM_{F}(T)$ and $EM_{Q}(T)$ we obtained the total $EM(T)$ of whole solar corona, including the average effect of both flare and quiescent plasma (see Fig.~\ref{fig:fqemd}). The quiescent component dominates at low temperatures, while the flaring $EM_{F}(T)$ dominates for $T\sim5-10$\,MK. At higher temperatures ($T\ge20$\,MK) $EM_{Q}(T)$ is larger than $EM_{F}(T)$. However results for such high temperatures must be considered with caution: because the two {\it Yohkoh}/SXT filters used for deriving $EM_{Q}(T)$ are not tailored for hot plasma \citep{OrlandoPeres2000}. In summary the total $EM(T)$ of the whole solar corona is double peaked, with the cooler peak, at $T\sim1-2$\,MK related to the quiescent corona, and the hotter peak, at $T\sim6-8$\,MK, to the flares.

\begin{figure}
\centering
\includegraphics[width=8.5cm]{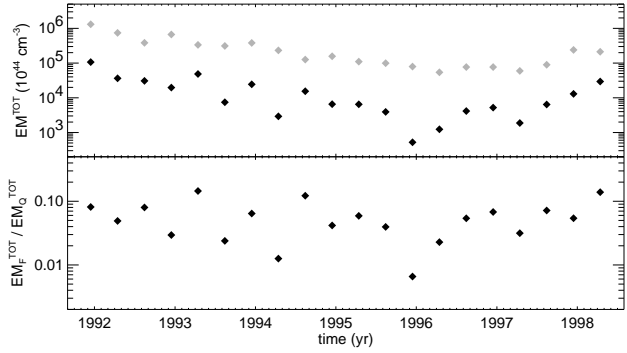} 
\caption{{\it Upper panel:} total emission measure of the flaring ($EM_{F}^{TOT}$, indicated in black) and quiescent ($EM_{Q}^{TOT}$, in gray) coronal plasma vs. time. {\it Lower panel:} ratio between total emission measure of flaring and coronal plasma vs. time.} 
\label{fig:emtotvstime}
\end{figure}

The variations during the solar cycle of the total amount of both quiescent and flaring emission measure (hereafter $EM_{Q}^{TOT}$ and $EM_{F}^{TOT}$) are shown in Fig.~\ref{fig:emtotvstime}. For both $EM_{Q}^{TOT}$ and $EM_{F}^{TOT}$ the observed uncertainties are much smaller than the size of the plotting symbols. In Fig.~\ref{fig:emtotvstime} both $EM_{Q}^{TOT}$ and $EM_{F}^{TOT}$ follow the solar cycle. The linear correlation coefficient of $EM_{Q}^{TOT}$ and $EM_{F}^{TOT}$ is 0.89, corresponding to a probability smaller than 0.1\% that the two quantities are uncorrelated. The variations of $EM_{Q}^{TOT}$ are smooth over time scales of a few years. Conversely $EM_{F}^{TOT}$, superimposed to the modulation due to the solar cycle, shows large variations also on time scales of a few months. However, irrespective of these short time scales fluctuations, the flare component at any time emerges over the quiescent emission measure distribution over the range $T\sim5-10$\,MK (see~Fig~\ref{fig:fqemd}).


We repeated the evaluation of the total emission measure of both flaring and quiescent plasma, $EM_{Q}^{HOT}$ and $EM_{F}^{HOT}$, summing only the $EM$ in the range $T\ge3$\,MK, a temperature range where both the {\it Yohkoh}/SXT and {\it GOES}/XRS are well sensitive (Fig.~\ref{fig:emhotvstime}). Again both components display a variation on time scales of a few years correlated with the solar cycle. They both show variations on shorter time scales (a few months). Their linear correlation coefficient is 0.94, corresponding to a probability $<0.1$\% that $EM_{Q}^{HOT}$ and $EM_{F}^{HOT}$ are uncorrelated.

The ratio between $EM_{Q}^{HOT}$ and $EM_{F}^{HOT}$ seems to be slightly larger near the solar minimum than during the solar maximum, suggesting that the relative contribution of flaring plasma with respect to the hot part of the quiescent corona tends to increase near the solar minimum.

\section{Discussion and conclusions}
\label{disc}

The main result of our study is that an unbiased and large sample of solar flares produces an emission measure distribution $EM_{F}(T)$, time-averaged over a period of one month, which peaks at $T\approx6-8$\,MK, and which is significantly larger than the quiescent emission measaure distribution for $T\sim5-10$\,MK. These properties of $EM_{F}(T)$ hold irrespective of the phase of the solar cycle.
 
\begin{figure}
\centering
\includegraphics[width=8.5cm]{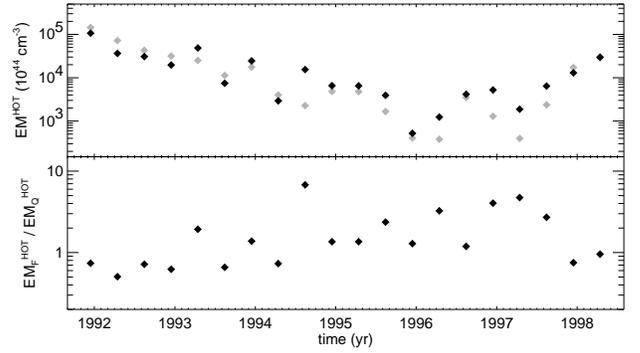} 
\caption{{\it Upper panel:} total emission measure for $T\ge3$\,MK of the flaring ($EM_{F}^{HOT}$, in black) and of the quiescent ($EM_{Q}^{HOT}$, in gray) coronal plasma vs. time. {\it Lower panel:} ratio between emission measure for $T\ge3$\,MK of flaring and coronal plasma vs. time.} 
\label{fig:emhotvstime}
\end{figure} 

Summing $EM_{F}(T)$ and $EM_{Q}(T)$ we measured for the first time the time-averaged $EM(T)$ of the whole solar corona, thus taking into account the average effect of solar flares and quiescent structures. The total solar $EM(T)$ is double peaked, with the cooler peak, at $T\sim1-2$\,MK due to the quiescent corona, and the hotter peak, at $T\sim6-8$\,MK, due to the flares.

In our study we also investigated how the quiescent and flare components of the solar corona vary during the solar cycle. The upper panel of Fig.~\ref{fig:emhotvstime} shows that the hot components ($T>3$\,MK) of the quiescent and flaring emission measure, $EM_{F}^{HOT}$ and $EM_{Q}^{HOT}$, vary similarly on time scales of few years following the solar cycle, but they also are strongly correlated on shorter time scales, i.e. a few months. It means that the amount of flaring plasma is highly correlated with the hot quiescent structures. The hot tail of quiescent emission measure distribution is due to the active regions and their cores \citep{OrlandoPeres2001}. So the observed correlation between flares and active regions may reflect that flares occur in active regions \citep[e.g.][]{BornmannShaw1994}.

The results obtained for the total $EM(T)$ provide new insight for the study of the solar corona heating. It has been proposed that the coronal heating is entirely due to flares, ranging from infrequent giant flares down to very frequent nanoflares \citep{Parker1988,Hudson1991,Klimchuk2006}. In this scenario both the quiescent and flare components of the solar corona are therefore related to flaring activity; $EM_{F}(T)$ is obviously due to the high energy tail of the flare distribution, while the plasma contributing to the $EM_{Q}(T)$ is instead due to less energetic flares (i.e., nano and microflares).

Generally a corona composed of identical loops heated by nanoflares has an $EM(T)$ characterized by one sharp peak \citep{Cargill1994,CargillKlimchuk2004}. The $EM(T)$ peak is approximately located at the temperature for which the radiative and conductive cooling time, $\tau_{r}$ and $\tau_{c}$, are equal. These two characteristic times, and hence the peak temperature, depend on the loop geometry and plasma density. Therefore a possible explanation of the double peaked $EM(T)$ of the solar corona is that two different loop distributions exist, as in the theoretical model explored by \citet{Guedel1997}, who obtained a double peaked $EM(T)$ by studying the effect of a power law flare distribution, $dN/dE\propto E^{-\alpha}$, on loops with different cross sections. However in the solar corona no evidence was found that major flares occur preferentially in a particular loop class.

\citet{ParentiBuchlin2006} on the other hand showed that the $EM(T)$ peak temperature can also be varied by changing the average flare energy. Therefore a double peaked $EM(T)$ can also arise from two different distributions for flares at different energy. This latter interpretation agrees with the idea that nanoflares and flares are not two portions of the same distribution.

In the present work we find that the solar total $EM(T)$ has two ever-present distinct peaks, one due to the quiescent corona and one to flares. This simple finding has a two-fold important implication: if we give credit to the hypothesis that the quiescent corona is heated by nanoflares, major flares and nanoflares belong to different populations; if we instead work in the assumption that flares and nanoflares belong to the same population, the implication is that nanoflares alone cannot explain the emission measure of the quiescent corona.


Beyond the understanding of coronal heating, the results obtained for the total $EM(T)$, flaring plus quiescent, provide new insight also for the link between the solar corona and the coronae of active stars. The emission measure distribution $EM(T)$ is a rather useful means to compare coronal plasmas. Active stellar coronae have an emission measure distribution with a large peak at $T\sim8-20$\,MK. Conversely in the solar corona significant amount of emission measure at these temperatures is observed only during flares, while quiescent structures have temperatures of a few MK. Starting from this consideration it has been proposed that the hot peak in the $EM(T)$ of active stellar coronae could be due to the presence of a large number of unresolved solar like flares \citep[e.g.][]{GuedelGuinan1997}.

The results for $EM_{F}(T)$ and $EM_{Q}(T)$ suggest some analogies with the coronae of active stars. Active stars display a double-peaked $EM(T)$, and we showed that the same happens for the solar corona. The solar $EM_{F}(T)$ is sharply peaked at temperatures (6-8\,MK) which are similar, albeit slightly cooler, to the temperatures of active coronae $EM(T)$ peak (8-20\,MK). Therefore it is more plausible that the hot peak of active coronae, in analogy with the solar case, is due to unresolved flares. Note that active stellar flares are on average hotter than the solar flares.

\begin{figure}
\centering
\includegraphics[width=8.5cm]{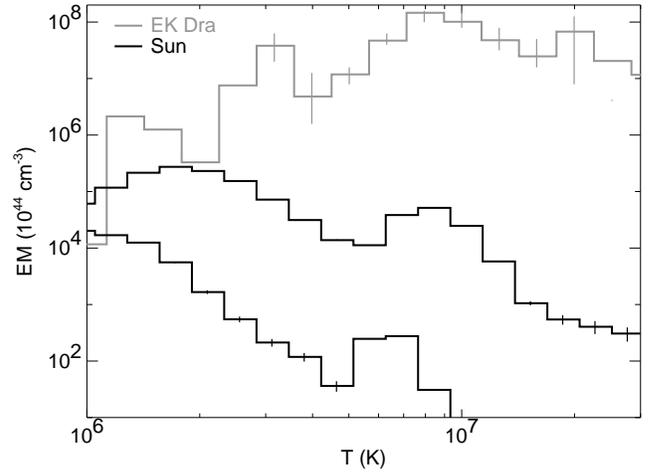} 
\caption{Solar emission measure distributions (flaring + quiescent) near the minimum and maximum phases of the solar cycle (December 1991 and December 1995, both in black) compared with the emission measure distribution of EK~Dra \citep[][, in gray]{ScelsiMaggio2005}. We show error bars in the EK~Dra distribution only for those temperature bins where the $EM$ values are constrained.}
\label{fig:compemd}
\end{figure}

As an example in Fig.~\ref{fig:compemd} the solar emission measure distribution, near the maximum and minimum phases of the solar cycle, is compared with the distribution of \object{EK~Dra} \citep{ScelsiMaggio2005}, which is a young solar analog: a $1.1\,M_{\sun}$ star just arrived on the main sequence, with $L_{X}\sim10^{30}\,{\rm erg\,s^{-1}}$. Inspecting the two solar $EM(T)$ note that the hot peak is located at higher temperature near the solar maximum: this is mainly due to the hot part of $EM_{Q}(T)$ which changes slope with the solar cycle (see Fig.~\ref{fig:fqemd}). More frequent solar-like flares would enhance $EM_{F}(T)$ to match the hot peak of the emission measure distribution of active stellar coronae\footnote{It is true if we assume that each flare occurs in a different coronal loop.}. On the other hand, in order for the solar plasma to match the $EM(T)$ of active stars, the $EM(T)$ of the quiescent plasma should undergo a smaller increment with respect to that of the flaring plasma. Then a different scaling of the quiescent and flaring plasma components would be required to explain the stellar activity.

In the strong assumption that the heating of the corona is entirely due to flares, we inferred that for the Sun either the flare distribution, or the confined plasma response to flares, or both, are bimodal. Comparing the solar $EM(T)$ with that of active stars, considering also how the $EM(T)$ shape changes with activity level, we inferred that the flare and quiescent components have different relative strengths at different stellar activity levels. This result implies that different parts of the flare distribution (i.e. flares, micro- and nano-flares) or the confined plasma response to flares and nano-flares, scale in different ways with stellar activity.

A possible scenario in this perspective is that of \citet{DrakePeres2000}, who argued that for increasing coronal activity, the large surface coverage of active regions and strong magnetic fields could provoke a change in the dissipation of magnetic energy, favoring more magnetic interactions and reconnections, and therefore a more efficient production of flares. However this scenario is not supported by the small filling factors inferred for the active stellar coronae \citep{TestaDrake2004b,NessGuedel2004}.

In conclusion, this work shows that taking into account both flaring and quiescent emission leads to bimodal emission measure distribution, yielding insight into stellar coronal physics; on the other hand this work gives a constraint onto the distributions of nano-, micro-, and macro-flares.

More precise estimations of the $EM(T)$ of the solar corona will be possible with the new instruments that will provide higher spatial resolution and better temperature estimation, for instance: the X-Ray Telescope (XRT) on board {\it Hinode} \citep{KosugiMatsuzaki2007}, or the Atmospheric Imaging Assembly (AIA) and the Extreme Ultraviolet Variability Experiment (EVE) both on board the forthcoming {\it Solar Dynamics Observatory} ({\it SDO}). {\it Hinode}/XRT collects X-ray images of the full disk Sun, with $1\,\arcsec$ angular resolution, in 9 different bands, allowing a very detailed thermal characterization of the whole corona and an excellent derivation of the coronal $EM(T)$. {\it SDO}/AIA will gather images of the whole solar corona in several narrow band filters with high spatial resolution ($1\,\arcsec$) and with high sampling rate ($\sim10$\,s). Thus {\it SDO}/AIA will allow to derive the whole coronal $EM(T)$ (flaring + quiescent) in a wide range of temperature with only one instrument. {\it SDO}/EVE will gather disk-integrated X-ray and UV spectra. Thus the derived solar $EM(T)$ can be compared with the stellar $EM(T)$, also obtained from disk-integrated spectra.


\appendix

\section{Quiescent contribution estimation in the GOES/XRS data}
\label{qflux}

For the correct analysis of {\it GOES}/XRS data it was important to individuate the level corresponding to the quiescent coronal emission in the one month-long {\it GOES}/XRS light curves in the soft and hard bands. Since the coronal quiescent emission may change over the period of one month, we divided the two {\it GOES}/XRS light curves in segments of 24\,h, and we assumed that the quiescent emission did not vary in each 24\,h sub-interval. We assumed that in each of these 24\,h sub-intervals the observed minimum values of the {\it GOES}/XRS fluxes is the quiescent emission. Since however the minimum fluxes of the soft and hard bands usually occurred at different times, we required that the soft and hard fluxes, representing the quiescent emission, should be simultaneous, and not significantly larger than their absolute minimum values. Therefore we adopted the following procedure: 1) we individuated the minimum flux in the hard band; 2) we selected all the bins for which the hard flux ranged between its minimum and its minimum increased by 10\%; 3) among these bins we selected that one (corresponding to the time $t_{\rm min}$) for which the soft fluxes is minimum; 4) the hard and soft fluxes recorded at $t_{\rm min}$ are considered representative of the quiescent emission for the 24\,h sub-interval. For each 24\,h-interval we subtracted the relevant quiescent fluxes from the soft and hard fluxes. This procedure for the search of the minimum was applied to the {\it GOES}/XRS light curves slightly smoothed (over a time lapse of $\sim30$\,s), to avoid problems due to low statistics.

\begin{acknowledgements}

CA, GP, SO, and FR acknowledge partial support for this work from contract ASI-INAF I/023/05/0 and from the Ministero dell'Universit\`a e della Ricerca.

\end{acknowledgements}

\bibliographystyle{aa} 
\bibliography{solaremd}

\end{document}